\documentclass[preprintnumbers,amsmath,amssymb,showpacs]{revtex4}

\usepackage[                   
            pdfstartview=FitH,
            colorlinks, 
            pdfborder=001,   
            linkcolor=blue,
            anchorcolor=blue,
            citecolor=blue,
            urlcolor=blue
            ]{hyperref}
\usepackage{graphicx}
\usepackage{dcolumn}
\usepackage{bm}
\usepackage{subfigure}
\usepackage{diagbox}
\usepackage[toc,page,title,titletoc,header]{appendix}
\begin{document}

\title{Efficient detection for quantum states containing fewer than $k$ unentangled particles in multipartite quantum systems}

\author{Yabin Xing}
 \affiliation {School of Mathematics and Science, Hebei GEO University, Shijiazhuang 050031,  China}

\author{Yan Hong}
 \affiliation {School of Mathematics and Science, Hebei GEO University, Shijiazhuang 050031,  China}

\author{Limin Gao}
 \affiliation {School of Mathematics and Science, Hebei GEO University, Shijiazhuang 050031,  China}

\author{Ting Gao}
\email{gaoting@hebtu.edu.cn} \affiliation {School of Mathematical Sciences, Hebei Normal University, Shijiazhuang 050024,  China}

\author{Fengli Yan}
\email{flyan@hebtu.edu.cn} \affiliation {College of Physics, Hebei Key Laboratory of Photophysics Research and Application,
Hebei Normal University, Shijiazhuang 050024,  China}

\begin{abstract}

In this paper, we mainly investigate the detection of quantum states containing  fewer than $k$ unentangled particles in multipartite quantum systems. Based on  calculations about operators, we derive two practical criteria for judging $N$-partite quantum states owning fewer than $k$ unentangled particles. In addition, we demonstrate the effectiveness of our frameworks through some concrete examples,
and specifically point out the quantum states  having fewer than $k$ unentangled particles that our methods can detect, while other criteria cannot recognize.

\end{abstract}


\maketitle

\section{Introduction}

Quantum entanglement \cite{EinsteinPodolsky35,Werner89} is a very special phenomenon in quantum systems and has been regarded as an important resource. Based on the properties of quantum entanglement, it can complete some tasks that cannot be completed by traditional methods, such as quantum communication and quantum computing \cite{RaussendorfBriegel2001}, quantum cryptography \cite{BB84,GisinRibordyTittelZbinden2002}, quantum dense coding \cite{HilleeryBuzek1999,BennettWiesner1992}, quantum teleportation \cite{PRL70.1895}.

In the theory of quantum entanglement, one of the most basic problems is to determine whether a quantum state is entangled or separable,  and great efforts have been made to try to solve this problem for a long time \cite{HofmannTakeuchi03,ChenAlbeverioFei05,Simon00,NhaZubairy08,SchwonnekDammeierWerner17,Rastegin15,GniewomirGiovanniDariusz20,Peres96,ChenWu03}.
In multipartite quantum systems, the entanglement of quantum states  is often characterized from different perspectives \cite{GuhneToth2009}, for example,  $k$-nonseparability is according to \textquotedblleft How many partitions are separable?\textquotedblright; $k$-partite entanglement is based on \textquotedblleft How many particles are entangled?\textquotedblright.
In the past few years, many entanglement criteria have been proposed in terms of tools \cite{WangWeiFei22,GabrielHiesmayrHuber10,HongLuoSong15,XuZhengZhuFei21,WangHou15,GaoYan2014,MaChen2010,HuberMintert2010,HyllusLaskowski12,EPJDHongGao,PRA82.062113,PRA86.062323,EPL104.20007}, which can only recognize certain multipartite entanglement, not all multipartite entanglement.
There is another way to depict entanglement  from  the number of unentangled  particles \cite{GuhneToth2009,
TothKnapp2009}, that is, the quantum states containing fewer than $k$ unentangled particles in $N$-partite quantum systems.  When the quantum states violate the criterion based on the variance \cite{Guhne2004},  the quantum states cannot be constructed by one or more unentangled particles. Using quantum Fisher information, any state that violates the framework   has fewer than $k$ unentangled particles \cite{Toth12}.
Another criterion gives an inequality that a quantum state  owning at least $k$ unentangled particles must satisfy \cite{HongQiGao21}.
The number of unentangled particles is closely related to metrological usefulness of quantum states \cite{Toth12},
so this prompts us to further explore these quantum states fewer than $k$ unentangled particles.

In this paper, we   further    research  the quantum states fewer than $k$ unentangled particles and  develop the detection frame of such  states.
In Section II, we   introduce some important concepts and symbols to be used in the following text. In Section III and Section IV,
we   present our main results, namely the detection methods of quantum states fewer than $k$ unentangled particles, and demonstrate the practicality and operability of our criteria through concrete examples, respectively.

\section{Preliminary}
In an $N$-partite quantum system with Hilber space $\mathcal{H}_1\otimes\mathcal{H}_2\otimes\cdots\otimes\mathcal{H}_N$, the pure state $|\psi\rangle$ contains $k$  unentangled particles if it can be written as $|\psi\rangle=\bigotimes\limits_{i=1}^{k+1}|\psi_{A_i}\rangle$,
where $A_1,A_2,\cdots,A_{k+1}$ constitutes a partition of $\{1,2,\cdots,N\}$, $|\psi_{A_i}\rangle$ is a single-partite state for $1\leq i\leq k$,  and $|\psi_{A_{k+1}}\rangle$ owns $N-k$ particles \cite{Toth12,
TothKnapp2009}.
If a mixed state $\rho$ can be represented as a mixture of the pure states containing $k$ or
more unentangled particles (may belong to different partitions), then $\rho$ contains at least $k$ unentangled particles \cite{Toth12,
TothKnapp2009}, otherwise the state $\rho$ has fewer than $k$ unentangled particles.

For Hilbert space  $\mathcal{H}_1\otimes\mathcal{H}_2\otimes\cdots\otimes\mathcal{H}_N$ with dim $\mathcal{H}_i=d_i, i=1,2,\cdots,N$, we first introduce the local permutation operator $P_\alpha$ and global permutation operator $P$,
\begin{equation}\label{}\nonumber
\begin{array}{rl}
P_\alpha(\bigotimes\limits_{i=1}^Nx_i)\otimes(\bigotimes\limits_{i=1}^Ny_i)=(\bigotimes\limits_{i\in\alpha}y_i\bigotimes\limits_{i\notin\alpha}x_i)\otimes
(\bigotimes\limits_{i\in\alpha}x_i\bigotimes\limits_{i\notin\alpha}y_i),
\end{array}
\end{equation}

\begin{equation}\label{}\nonumber
\begin{array}{rl}
P(\bigotimes\limits_{i=1}^Nx_i)\otimes(\bigotimes\limits_{i=1}^Ny_i)=(\bigotimes\limits_{i=1}^Ny_i)\otimes(\bigotimes\limits_{i=1}^Nx_i),
\end{array}
\end{equation}
where  $\alpha$ is any subset of $\{1,2,\cdots,N\}$, $\bigotimes\limits_{i=1}^Nx_i$ and  $\bigotimes\limits_{i=1}^Ny_i$  are any operators of Hilbert space  $\mathcal{H}_1\otimes\mathcal{H}_2\otimes\cdots\otimes\mathcal{H}_N$ with $x_i$ and $y_i$ acting on $\mathcal{H}_i$. In particular, when $\alpha$ is taken as  $\{1,2,\cdots,N\}$ or $\{i\}$ with $1\leq i\leq N$, $P_\alpha$ is abbreviated as $P$, $P_i$, respectively.

\section{Detection of  quantum states containing fewer than $k$ unentangled particles}
Now, we state our main results.

\emph{Theorem 1.} In Hilbert space  $\mathcal{H}_1\otimes\mathcal{H}_2\otimes\cdots\otimes\mathcal{H}_N$ with dim $\mathcal{H}_i=d_i, i=1,2,\cdots,N$, if a quantum state $\rho$ contains at least $k$ unentangled particles for $1\leqslant k\leqslant N-1$, then we have
\begin{equation}\label{kunentangled4}
\begin{array}{rl}
(2^{k+1}-2)\sqrt{\textrm{Tr}\big[(X^\dagger\otimes Y^\dagger)\rho^{\otimes2}P(X\otimes Y)\big]}
\leq\sum\limits_{\{\alpha\}}\sqrt{\textrm{Tr}\big[(X^\dagger\otimes Y^\dagger)P^\dagger_\alpha\rho^{\otimes2}P_\alpha(X\otimes Y)\big]},
\end{array}
\end{equation}
where $\{\alpha\}$  consists of  nonempty proper subsets of $\{1,2,\cdots,N\}$,  $X=\bigotimes\limits_{i=1}^Nx_i$ and  $Y=\bigotimes\limits_{i=1}^Ny_i$  are any operators of Hilbert space  $\mathcal{H}_1\otimes\mathcal{H}_2\otimes\cdots\otimes\mathcal{H}_N$. If a quantum state $\rho$ violates inequality (\ref{kunentangled4}), then it contains  fewer than $k$ unentangled particles.

\emph{Proof.}
First we prove that  the inequality (\ref{kunentangled4}) is true for any pure state. Suppose that pure state $\rho=|\psi\rangle\langle\psi|$
contains at least $k$ unentangled particles, then there exits a partition $A_1|A_2|\cdots|A_{k+1}$ such that
 $|\psi\rangle=\bigotimes\limits_{m=1}^{k+1}|\varphi_{A_m}\rangle$.
By Cauchy-Schwarz inequality, we have
\begin{equation}\label{Th1.1}
\begin{array}{rl}
&\textrm{Tr}\big[(X^\dagger\otimes Y^\dagger)\rho^{\otimes2}P(X\otimes Y)\big]
=|\langle\psi|YX^\dag|\psi\rangle|^2\leq\langle\psi|YY^\dag|\psi\rangle\langle\psi|XX^\dag|\psi\rangle.
\end{array}
\end{equation}
For the partition $A_1|A_2|\cdots|A_{k+1}$, let $\alpha_{j_1,\cdots,j_n}$ be a set of any $n$ subsets $A_{j_t}$ (that is, $\alpha_{j_1,\cdots,j_n}=A_{j_1}\bigcup \cdots\bigcup A_{j_n}$), $\overline{\alpha}_{j_1,\cdots,j_n}$ be complement
($\overline{\alpha}_{j_1,\cdots,j_n}=\{1,2,\cdots,N\}-\alpha_{j_1,\cdots,j_n}=A_{j_{n+1}}\bigcup \cdots\bigcup A_{j_{k+1}}$).
Thus we get
\begin{equation}\label{Th1.2}
\begin{array}{rl}
&\textrm{Tr}\big[(X^\dagger\otimes Y^\dagger)P_{\alpha_{j_1,j_2,\cdots,j_n}}^\dagger\rho^{\otimes2}P_{\alpha_{j_1,j_2,\cdots,j_n}}(X\otimes Y)\big]\\
=&\textrm{Tr}\big\{[(\bigotimes\limits_{i\in\alpha_{j_1,j_2,\cdots,j_n}}y_{i}^\dagger)\otimes(\bigotimes\limits_{i\notin\alpha_{j_1,j_2,\cdots,j_n}}x_{i}^\dagger)](\bigotimes\limits_{m=1}^{k+1}|\varphi_{A_m}\rangle\langle\varphi_{A_m}|)
[(\bigotimes\limits_{i\in\alpha_{j_1,j_2,\cdots,j_n}}y_{i})\otimes(\bigotimes\limits_{i\notin\alpha_{j_1,j_2,\cdots,j_n}}x_{i})]\big\}\\
&
\times\textrm{Tr}\big\{[(\bigotimes\limits_{i\in\alpha_{j_1,j_2,\cdots,j_n}}x_{i}^\dagger)\otimes(\bigotimes\limits_{i\notin\alpha_{j_1,j_2,\cdots,j_n}}y_{i}^\dagger)]
(\bigotimes\limits_{m=1}^{k+1}|\varphi_{A_m}\rangle\langle\varphi_{A_m}|)[(\bigotimes\limits_{i\in\alpha_{j_1,j_2,\cdots,j_n}}x_{i})\otimes(\bigotimes\limits_{i\notin\alpha_{j_1,j_2,\cdots,j_n}}y_{i})]\big\}\\
=&\Big(\prod\limits_{t=1}^n\langle\varphi_{A_{j_t}}|\bigotimes\limits_{i\in A_{j_t}}y_iy_i^\dag|\varphi_{A_{j_t}}\rangle\prod\limits_{t=n+1}^{k+1}\langle\varphi_{A_{j_t}}|\bigotimes\limits_{i\in A_{j_t}}x_ix_i^\dag|\varphi_{A_{j_t}}\rangle\Big)
\times\Big(\prod\limits_{t=1}^n\langle\varphi_{A_{j_t}}|\bigotimes\limits_{i\in A_{j_t}}x_ix_i^\dag|\varphi_{A_{j_t}}\rangle\prod\limits_{t=n+1}^{k+1}\langle\varphi_{A_{j_t}}|\bigotimes\limits_{i\in A_{j_t}}y_iy_i^\dag|\varphi_{A_{j_t}}\rangle\Big)\\
=&\langle\psi|YY^\dag|\psi\rangle\langle\psi|XX^\dag|\psi\rangle.
\end{array}
\end{equation}
Eq. (\ref{Th1.1}) and Eq. (\ref{Th1.2}) ensure that the inequality (\ref{kunentangled4}) holds for any pure state containing at least $k$ unentangled particles.

Let  $\rho$ be  a  mixed state containing at least $k$ unentangled particles, then it can be rewritten as
$\rho=\sum\limits_ip_i|\varphi_i\rangle\langle\varphi_i|$  with the pure state $\rho_i=|\varphi_i\rangle\langle\varphi_i|$ containing at least $k$ unentangled particles. Thus we have
\begin{align}
&(2^{k+1}-2)\sqrt{\textrm{Tr}\big[(X^\dagger\otimes Y^\dagger)\rho^{\otimes2}P(X\otimes Y)\big]}\label{}\nonumber\\
\leq&(2^{k+1}-2)\sum\limits_ip_i\sqrt{\textrm{Tr}(X^\dagger\otimes Y^\dagger)\rho_i^{\otimes2}P(X\otimes Y)}\label{1.1}\\
\leq&\sum\limits_ip_i\sum\limits_{\{\alpha\}}\sqrt{\textrm{Tr}(X^\dagger\otimes Y^\dagger)P_{\alpha}^\dagger\rho_i^{\otimes2}P_{\alpha}(X\otimes Y)}\label{1.2}\\
\leq&\sum\limits_{\{\alpha\}}\sqrt{\Big\{\sum\limits_ip_{i}\textrm{Tr}[(\bigotimes\limits_{i\in\alpha}y_{i}^\dagger)\otimes(\bigotimes\limits_{i\notin\alpha}x_{i}^\dagger)
\rho_i(\bigotimes\limits_{i\in\alpha}y_{i})\otimes(\bigotimes\limits_{i\notin\alpha}x_{i})]\Big\}\Big\{\sum\limits_ip_{i}\textrm{Tr}
[(\bigotimes\limits_{i\in\alpha}x_{i}^\dagger)\otimes(\bigotimes\limits_{i\notin\alpha}y_{i}^\dagger)\rho_i
(\bigotimes\limits_{i\in\alpha}x_{i})\otimes(\bigotimes\limits_{i\notin\alpha}y_{i})]\Big\}}\label{1.3}\\
=&\sum\limits_{\{\alpha\}}\sqrt{\textrm{Tr}\big[(X^\dagger\otimes Y^\dagger)P^\dagger_\alpha\rho^{\otimes2}P_\alpha(X\otimes Y)\big]}.\label{}\nonumber
\end{align}
Here we have used  triangle inequality, validity of inequality (\ref{kunentangled4}) for any pure state  containing at least $k$ unentangled particles, and Cauchy-Schwarz inequality at the inequalities (\ref{1.1}), (\ref{1.2}) and (\ref{1.3}), respectively.
The above proof is exactly what we want.

\emph{Theorem 2.} In a Hilbert space  $\mathcal{H}^{\otimes N}=\mathcal{H}_1\otimes\cdots\otimes \mathcal{H}_N$ with dim $\mathcal{H}=d$,
 any $N$-partite quantum state $\rho$  containing at least $k$ unentangled particles must  satisfy the following inequality,
\begin{equation}\label{kunentangled5}
\begin{array}{rl}
&\sum\limits_{\substack{s,t\in\omega\\1\leq i\neq j\leq N}}\sqrt{\textrm{Tr}\big[({X_{i}^{s}}^\dagger\otimes {X_{j}^{t}}^\dagger)\rho^{\otimes2}P(X_{i}^{s}\otimes X_{j}^{t})\big]}\\
\leq & \sum\limits_{\substack{s,t\in\omega\\1\leq i\neq j\leq N}}\sqrt{\textrm{Tr}\big[({X_{i}^{s}}^\dagger\otimes {X_{j}^{t}}^\dagger)P^\dagger_i\rho^{\otimes2}P_i(X_{i}^{s}\otimes X_{j}^{t})\big]}
+T(N-k-1)\sum\limits_{\substack{s\in \omega\\1\leq i\leq N}}\sqrt{\textrm{Tr}\big[({X_{i}^{s}}^\dagger\otimes {X_{i}^{s}}^\dagger)\rho^{\otimes2}(X_{i}^{s}\otimes X_{i}^{s})\big]}
\end{array}
\end{equation}
for $2\leq k\leq N-1$, and
\begin{equation}\label{kunentangled6}
\begin{array}{rl}
\textrm{Tr}\big[({X_{i}^{s}}^\dagger\otimes {X_{j}^{t}}^\dagger)\rho^{\otimes2}P(X_{i}^{s}\otimes X_{j}^{t})\big]
\leq\textrm{Tr}\big[({X_{i}^{s}}^\dagger\otimes {X_{j}^{t}}^\dagger)P^\dagger_i\rho^{\otimes2}P_i(X_{i}^{s}\otimes X_{j}^{t})\big]
\end{array}
\end{equation}
for $k=1$.
Here $\omega=\{\omega_1,\cdots,\omega_T\}$ is a set of arbitrary $T$ operators in $\mathcal{H}$,   $X=\bigotimes\limits_{i=1}^Nx_i$  are any operator with $x_i$ acting on subsystem $\mathcal{H}_i$,
 $X_{i}^{s}=(\bigotimes\limits_{m=1}^{i-1}x_m)\otimes \omega_s \otimes (\bigotimes\limits_{m=i+1}^{N}x_m)$ is the  operator with $\omega_s$ acting on subsystem $\mathcal{H}_i$ and $x_m$ acting on subsystem $\mathcal{H}_m$ for $m\neq i$.
If a quantum state $\rho$ does not satisfy the above inequality,  it contains  fewer than $k$ unentangled particles.

\emph{Proof}.
 Suppose that the quantum state $\rho=|\psi\rangle\langle\psi|$ is pure state where $|\psi\rangle=\bigotimes\limits_{m=1}^{k+1}|\varphi_{A_m}\rangle$ contains at least $k$ unentangled particles under the partition $A_1|\cdots|A_{k+1}$ with the subset $A_{k+1}$ owning $N-k$ particles and  each of rest subsets $A_m$ owning 1 particle.
After calculations, we can easily obtain
\begin{equation}\label{}\nonumber
\begin{array}{rl}
&\sqrt{\textrm{Tr}\big[({X_{i}^{s}}^\dagger\otimes {X_{j}^{t}}^\dagger)\rho^{\otimes2}P(X_{i}^{s}\otimes X_{j}^{t})\big]}\\
=&|\langle\psi| X_{j}^{t}{X_{i}^{s}}^\dagger|\psi\rangle|\\
\leq&\sqrt{\langle\psi| X_{j}^{t}{X_{j}^{t}}^\dagger|\psi\rangle\langle\psi| X_{i}^{s} {X_{i}^{s}}^\dagger|\psi\rangle}\\
\leq
&\dfrac{\sqrt{\textrm{Tr}\big[({X_{j}^{t}}^\dagger\otimes {X_{j}^{t}}^\dagger)\rho^{\otimes2}(X_{j}^{t}\otimes X_{j}^{t})\big]}+\sqrt{\textrm{Tr}\big[({X_{i}^{s}}^\dagger\otimes{X_{i}^{s}}^\dagger)\rho^{\otimes2}(X_{i}^{s}\otimes
X_{i}^{s})\big]}}{2},
\end{array}
\end{equation}
when $i$ and $j$ are both in  the subset $A_{k+1}$, where these two inequalities are true by using Cauchy-Schwarz inequality and the mean inequality;
and
\begin{equation}\label{}\nonumber
\begin{array}{rl}
&\sqrt{\textrm{Tr}\big[({X_{i}^{s}}^\dagger\otimes {X_{j}^{t}}^\dagger)\rho^{\otimes2}P(X_{i}^{s}\otimes X_{j}^{t})\big]}\\
=&\Big|\langle\psi|X_{j}^{t}{X_{i}^{s}}^\dagger|\psi\rangle\Big|\\
\leq&\sqrt{\textrm{Tr}\big[({X_{i}^{s}}^\dagger\otimes {X_{j}^{t}}^\dagger)P^\dagger_i\rho^{\otimes2}P_i(X_{i}^{s}\otimes X_{j}^{t})\big]}
\end{array}
\end{equation}
when $i$  and $j$ belong to different subsets $A_{l}$, $A_{l'}$.
Based on the above two cases, so we can get
\begin{equation}\label{}\nonumber
\begin{array}{rl}
&\sum\limits_{\substack{s,t\in\omega\\1\leq i\neq j\leq N}}\sqrt{\textrm{Tr}\big[({X_{i}^{s}}^\dagger\otimes {X_{j}^{t}}^\dagger)\rho^{\otimes2}P(X_{i}^{s}\otimes X_{j}^{t})\big]}\\
=&\sum\limits_{\substack{s,t\in\omega\\i,j\in{A_{k+1}},i\neq j}}\sqrt{\textrm{Tr}\big[({X_{i}^{s}}^\dagger\otimes{X_{j}^{t}}^\dagger)\rho^{\otimes2}P(X_{i}^{s}\otimes X_{j}^{t})\big]}+\sum\limits_{\substack{s,t\in\omega\\i\in{A_l},j\in{A_{l'}},l\neq l'}}
\sqrt{\textrm{Tr}\big[({X_{i}^{s}}^\dagger\otimes{X_{j}^{t}}^\dagger)\rho^{\otimes2}P(X_{i}^{s}\otimes
X_{j}^{t})\big]}\\
\leq&\sum\limits_{\substack{s,t\in\omega\\i,j\in{A_{k+1}},i\neq j}}
\dfrac{\sqrt{\textrm{Tr}\big[({X_{j}^{t}}^\dagger\otimes{X_{j}^{t}}^\dagger)P^\dagger_j\rho^{\otimes2}P_j(X_{j}^{t}\otimes
X_{j}^{t})\big]}+\sqrt{\textrm{Tr}\big[({X_{i}^{s}}^\dagger\otimes{X_{i}^{s}}^\dagger)P^\dagger_i\rho^{\otimes2}P_i(X_{i}^{s}\otimes
X_{i}^{s})\big]}}{2}\\
&+\sum\limits_{\substack{s,t\in\omega\\i\in{A_l},j\in{A_{l'}},l\neq l'}}
\sqrt{\textrm{Tr}\big[({X_{i}^{s}}^\dagger\otimes{X_{j}^{t}}^\dagger)P^\dagger_i\rho^{\otimes2}P_i(X_{i}^{s}\otimes X_{j}^{t})\big]}\\
\leq&\sum\limits_{\substack{s,t\in\omega\\1\leq i\neq j\leq N}}\sqrt{\textrm{Tr}\big[({X_{i}^{s}}^\dagger\otimes {X_{j}^{t}}^\dagger)P^\dagger_i\rho^{\otimes2}P_i(X_{i}^{s}\otimes X_{j}^{t})\big]}
+T(N-k-1)\sum\limits_{\substack{s\in \omega\\1\leq i\leq N}}\sqrt{\textrm{Tr}\big[({X_{i}^{s}}^\dagger\otimes {X_{i}^{s}}^\dagger)\rho^{\otimes2}(X_{i}^{s}\otimes X_{i}^{s})\big]}.
\end{array}
\end{equation}
This shows that the inequality (\ref{kunentangled5}) holds for any pure state containing at least $k$ unentangled particles.

 Next, let $\rho=\sum\limits_{m}p_{m}\rho_{m}=\sum\limits_{m}p_{m}|\varphi_{A_m}\rangle\langle\varphi_{A_m}|$ be a mixed state with the pure state $\rho_{m}=|\varphi_{A_m}\rangle\langle\varphi_{A_m}|$ containing at least $k$ unentangled particles, then we have
\begin{align}
&\sum\limits_{\substack{s,t\in\omega\\1\leq i\neq j\leq N}}\sqrt{\textrm{Tr}\big[({X_{i}^{s}}^\dagger\otimes {X_{j}^{t}}^\dagger)\rho^{\otimes2}P(X_{i}^{s}\otimes X_{j}^{t})\big]}\nonumber\\
\leq&\sum\limits_mp_m\sum\limits_{\substack{s,t\in\omega\\1\leq i\neq j\leq N}}\sqrt{\textrm{Tr}\big[({X_{i}^{s}}^\dagger\otimes{X_{j}^{t}}^\dagger)\rho_{m}^{\otimes2}P(X_{i}^{s}\otimes X_{j}^{t})\big]}\label{2.1}\\
\leq&\sum\limits_mp_{m}\sum\limits_{\substack{s,t\in\omega\\1\leq i\neq j\leq N}}\sqrt{\textrm{Tr}\big[({X_{i}^{s}}^\dagger\otimes{X_{j}^{t}}^\dagger)P_{i}^\dagger
\rho_{m}^{\otimes2}P_{i}(X_{i}^{s}\otimes X_{j}^{t})\big]}\nonumber\\
&+T(N-k-1)\sum\limits_mp_m\sum\limits_{\substack{s\in \omega\\
1\leq i\leq N}}\sqrt{\textrm{Tr}\big[({X_{i}^{s}}^\dagger\otimes{X_{i}^{s}}^\dagger)
\rho_{m}^{\otimes2}(X_{i}^{s}\otimes X_{i}^{s})\big]}\label{2.2}\\
\leq&\sum\limits_{\substack{s,t\in\omega\\1\leq i\neq j\leq N}}\sqrt{\big[\sum\limits_mp_m\textrm{Tr}(X^\dagger\rho_{m}X)\big]\big[\sum\limits_mp_m\textrm{Tr}((X^{st}_{ij})^\dagger\rho_{m}X^{st}_{ij})\big]}
+T(N-k-1)\sum\limits_{\substack{s\in \omega\\1\leq i\leq N}}\textrm{Tr}\big[(X^{s}_{i})^\dagger\sum\limits_mp_m\rho_{m}X^{s}_{i}\big]\label{2.3}\\
=&\sum\limits_{\substack{s,t\in\omega\\1\leq i\neq j\leq N}}\sqrt{\textrm{Tr}\big[({X_{i}^{s}}^\dagger\otimes {X_{j}^{t}}^\dagger)P^\dagger_i\rho^{\otimes2}P_i(X_{i}^{s}\otimes X_{j}^{t})\big]}
+T(N-k-1)\sum\limits_{\substack{s\in \omega\\1\leq i\leq N}}\sqrt{\textrm{Tr}\big[({X_{i}^{s}}^\dagger\otimes {X_{i}^{s}}^\dagger)\rho^{\otimes2}(X_{i}^{s}\otimes X_{i}^{s})\big]},\nonumber
\end{align}
where  $X_{ij}^{st}=(\bigotimes\limits_{m\neq i \textrm{ and } m\neq j}x_m)\otimes w_s\otimes w_t$ with $x_m$, $w_s$ and  $w_t$ acting on subsystem $\mathcal{H}_m$, $\mathcal{H}_i$ and $\mathcal{H}_j$, respectively. The inequality (\ref{2.1}), (\ref{2.2}) and (\ref{2.3}) are established by triangle inequality, validity of inequality (\ref{kunentangled5}) for any pure state  containing at least $k$ unentangled particles, and Cauchy-Schwarz inequality, respectively.
This shows that the inequality (\ref{kunentangled5})  holds for any mixed state containing at least $k$ unentangled particles.
We can similarly prove inequality (\ref{kunentangled6}).

\section{Illustration}
In this section,  we will demonstrate the operability and efficiency of our framework by applying it on typical quantum states.
It is worth noting that our criteria has better detection performance in the following two explicit examples.

\emph{Example 1.} Consider the family of  $8$-qubit quantum states
$\rho(p)=p|G_{8}\rangle\langle G_{8}|+\dfrac{1-p}{2^{8}}\textbf{1},$
with $|G_{8}\rangle=\dfrac{|0\rangle^{\otimes 8}+|1\rangle^{\otimes 8}}{\sqrt{2}}$ being $8$-qubit GHZ state.

These quantum states $\rho(p)$  contain fewer than $k$ unentangled particles when $p_k<p\leq p_k'$, which only
can be  detected by our Theorem 1 with choosing $x_{i}=|1\rangle\langle0|$, $y_{i}=|0\rangle\langle0|$, but not by observation 5 in Ref.\cite{Toth12}. This indicates that our Theorem 1 has  more efficient detection than observation 5 in Ref.\cite{Toth12} for the family of quantum states $\rho(p)$. The specific values of $p_k$ and $p'_k$ are shown in Table I.

\begin{table}
\caption{\label{tab:table1}The
thresholds of the quantum states   containing fewer than $k$ unentangled particles  for $\rho(p)=p|G_{8}\rangle\langle G_{8}|+\dfrac{1-p}{2^{8}}\textbf{1}$.  When $p_{k}<p \leq 1$ and ${p_k}'<p \leq 1$, $\rho(p)$ contains fewer than $k$ unentangled particles  captured by our Theorem 1 and observation 5 in Ref.\cite{Toth12}, respectively. }
\begin{ruledtabular}
\begin{tabular}{cccccccc}
\diagbox [width=2em,trim=l]{$p$}{$k$}&1&2&3&4&5&6&7\\
\hline
$p_{k}$ & 0.4980 & 0.2485 & 0.1241 & 0.0620 & 0.0310 & 0.0155 & 0.0078 \\
  ${p_k}'$ & 0.8015 & 0.6279 & 0.4790 & 0.3550  & 0.2557 & 0.1811  & 0.1315 \\
\end{tabular}
\end{ruledtabular}
\end{table}

\emph{Example 2.} Considering the family of $N$-partite quantum states,
\begin{eqnarray*}
\rho(p,q)=p|W\rangle\langle W|+q|\widetilde{W}\rangle\langle\widetilde{W}|+\dfrac{1-p-q}{d^{N}}\textbf{1},
\end{eqnarray*}
where $d\geq3$, $|W\rangle=\dfrac{1}{\sqrt{N(d-1)}}\sum\limits_{i=1}^{d-1}(|0\cdots00i\rangle+|0\cdots0i0\rangle+\cdots+|i0\cdots00\rangle)$ and $|\widetilde{W}\rangle=\sigma^{\otimes N}|W\rangle$ with $\sigma|0\rangle=|1\rangle, \cdots, \sigma|d-2\rangle=|d-1\rangle, \sigma|d-1\rangle=|0\rangle$.

Choose $x_{i}=|0\rangle\langle0|$ and $\{\omega_1,\cdots,\omega_T\}=\{|1\rangle\langle0|,|2\rangle\langle0|,\cdots,|d-1\rangle\langle0|\}$  (or $x_{i}=|1\rangle\langle1|$ and  $\{\omega_1,\cdots,\omega_T\}=\{|0\rangle\langle1|,|0\rangle\langle2|,\cdots,|0\rangle\langle d-1|\}$),
our Theorem 2 can always detect some quantum states containing fewer than $k$ unentangled particles.
Since the observation 5 of Ref.\cite{Toth12} only works for $d=2$, it can't recognize any quantum states containing fewer than $k$ unentangled particles for $d\geq 3$. This indicates that our Theorem 2 is more powerful than observation 5 of Ref.\cite{Toth12} for $N$-partite quantum states $\rho(p,q)$.
When $N=5, d=4$, we describe the specific ranges of quantum states containing fewer than $k$ unentangled particles for $k=1,2,3,4$ in Fig. 1.

\begin{figure}
\begin{center}
{\includegraphics[scale=0.6]{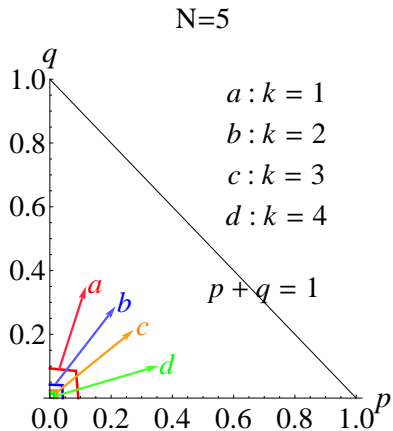}} \caption[Illustration of
]{(Color online) Detection power of Theorem 2 for $\rho(p,q)=p|W\rangle\langle W|+q|\widetilde{W}\rangle\langle\widetilde{W}|+\dfrac{1-p-q}{d^{N}}\textbf{1}$ for $k=1,2,3,4$ when $N=5,d=4$. The area enclosed by red line $a$ ( blue line $b$, orange line $c$, green line $d$ ), the $p$ axis, the $q$ axis and line $p+q=1$ represents quantum states containing  fewer than 1 (2, 3, 4) unentangled particles, respectively.}
\end{center}
\end{figure}

In particular, for $\rho(p,0)=p|W\rangle\langle W|+\dfrac{1-p}{d^{N}}\textbf{1}$, the quantum states $\rho(p,0)$ contain fewer than $k$ unentangled particles when $$p>\dfrac{N(d-1)(2N-k-2)}{kd^N+N(d-1)(2N-k-2)}.$$
Let $p_{(N,k,d)}:=\dfrac{N(d-1)(2N-k-2)}{kd^N+N(d-1)(2N-k-2)} $. For any $d$, when $1\leq k\leq N$, the larger $k$ is,  the smaller $p_{(N,k,d)}$ is.
For any $d$, it's easy to see that $\lim\limits_{N\rightarrow+\infty}p_{(N,k,d)}=0$.
This means that as $N$ increases, our Theorem 2  can identify more and more quantum states containing fewer than $k$ ($1\leq k\leq N$) unentangled particles for any $d$.
For any $N$ and $1\leq k\leq N$, it is obvious that $\lim\limits_{d\rightarrow+\infty}p_{(N,k,d)}=0$. This implies that as $d$ increases, our Theorem 2  can detect more and more quantum states containing fewer than $k$ unentangled particles.

\section{Conclusions}
In this work, we propose two practical  methods  for the detection of quantum states containing fewer than $k$ unentangled particles based on some permutations and  operators. These  methods are practical in two senses: first, our methods don't  involve optimization problems, only require some algebraic operations, this demonstrates their operability; second, our methods  can  identify some quantum states containing fewer than $k$ unentangled particles that are not recognized by other criteria, this shows our methods can be efficiently applied in practice.
As a consequence, the two criteria have good application potential for the detection of quantum states containing fewer than $k$ unentangled particles  in multipartite quantum systems.

\begin{center}
{\bf ACKNOWLEDGMENTS}
\end{center}

This work was supported by  the National Natural Science Foundation of China under Grant Nos. 62271189, 12071110, 11701135,  funded by Science and Technology Project of Hebei Education Department under Grant No.  ZD2021066, supported by National Pre-research Funds of Hebei GEO University in 2023 (Grant KY202316), PhD Research Startup Foundation of Hebei GEO University (Grant BQ201615).

\end{document}